\documentstyle[11pt,ws11,epsf]{article}

\def\LQCD{\Lambda_{\rm QCD}}
\def\mps{\Delta m_{\mbox{\scriptsize 1P--1S}}}
\def\MSbar{{\overline{\rm MS}}}
\newcommand{\half}{\mbox{\small $\frac{1}{2}$}}

\newcommand{\cH}{{\cal H}}
\newcommand{\cO}{{\cal O}}

\def\dfrac#1#2{{\displaystyle \frac{#1}{#2}}}

\preprint{FERMILAB-CONF-95/067-T \\ {\tt hep-ph/9504249}}
\title{\bf LATTICE QCD AND THE STANDARD MODEL%
\thanks{Invited paper presented at 
{\em Perspectives in Particle Physics '94}, 
the Seventh Adriatic Meeting on Particle Physics, Brijuni, Croatia, 
13--20 September 1994} 
}
\author{ANDREAS S. KRONFELD \\[0.3cm]
{\it Theoretical Physics Group,
Fermi National Accelerator Laboratory,} \\
{\it P.O. Box 500, Batavia, IL 60510, USA} }
\abstract{Most of the poorly known parameters of the Standard Model
cannot be determined without reliable calculations in nonperturbative
QCD.
Lattice gauge theory provides a first-principles definition of the
required functional integrals, and hence offers ways of performing
these calculations.
This paper reviews the progress in computing hadron spectra and
electroweak matrix elements needed to determine $\alpha_S$, the quark
masses, and the Cabibbo-Kobayashi-Maskawa matrix.}
\begin{document}
\setcounter{page}{0} 
\maketitle
\vfill \newpage 
\section{Introduction}
Many contemporary reviews of elementary particle physics start by
celebrating (or lamenting!) the success of the Standard Model.
Indeed, with some nineteen%
\footnote{I assume the neutrinos are massless, but count the ``vacuum
angle'' of QCD.}
parameters the SU(3)$\times$SU(1)$\times$U(1) gauge theory explains an
enormous array of experiments.
Even a terse compendium\cite{PDG94} of the experiments is more than big
enough to fill a phone book.
A glance at Table~\ref{table:sm} shows, however, that roughly half of
the parameters are not so well determined.
To test the Standard Model stringently, and thus to gain an inkling of
what lies beyond, we must learn the values of these parameters more
precisely.
\begin{table}
\caption[table:sm]{Parameters of the standard model and lattice
calculations that will help determine them.
Numerical values taken from the 1994 Review of Particle
Properties,\cite{PDG94} except $m_t$ (\refcite{CDF94}), $\sin\delta$,
and $\theta_{\rm QCD}$.  The strong coupling $\alpha_S$ refers to the
$\MSbar$ scheme at $M_Z$.
Adapted from \refcite{Kro93}.}\label{table:sm}
\begin{center} \begin{tabular}{c@{\hspace{2.0em}}c@{\hspace{2.0em}}c}
\hline \hline
\hspace{1em}parameter & value or range & related lattice calculations   \\
\hline
\multicolumn{3}{l}{\em gauge couplings}\\
  $\alpha_{\rm em}$  &   $1/137.036$    &                 \\
     $10^5G_F$       & 1.166 GeV$^{-2}$ &                 \\
     $\alpha_S$      &  $0.116\pm0.005$ &
$\Delta m_{\mbox{\scriptsize 1P--1S}}$; scaling \\
\multicolumn{3}{l}{\em electroweak masses}\\
$m_Z$                &   91.19 GeV      &        \\
$m_H$                &   $>58$ GeV      &        \\
\multicolumn{3}{l}{\em lepton masses}\\
$m_e$                &   0.51100 MeV    &        \\
$m_\mu$              &    105.66 MeV    &        \\
$m_\tau$             &    1777 MeV      &        \\
\multicolumn{3}{l}{\em quark masses}\\
$m_u$                &     2--8   MeV   & $m_\pi^2$, $m_K^2$ \\
$m_d$                &     5--15  MeV   & $m_\pi^2$, $m_K^2$ \\
$m_s$                &   100--300 MeV   & $m_K^2$ \\
$m_c$                &   1.0--1.6 GeV   & $m_{J/\psi}$   \\
$m_b$                &   4.1--4.5 GeV   & $m_{\Upsilon}$ \\
$m_t$                & $174\pm10^{+13}_{-12}$ GeV &      \\
\multicolumn{3}{l}{\em CKM matrix}\\
$s_{12}$             & 0.218--0.224     & $K\to \pi e\nu$ \\
$s_{23}$             & 0.032--0.048     & $B\to D^* l\nu$ \\
$s_{13}$             & 0.002--0.005     & $B\to \pi l\nu$ \\
$\sin \delta$        &    $\neq 0$      & $B_K$, $B_B$, $B_{B_s}$ \\
\multicolumn{3}{l}{\em QCD vacuum angle}\\
$\theta_{\rm QCD}$   &   $<10^{-9}$     & $d_n$   \\
\hline \hline
\end{tabular} \end{center}\end{table}

Except for the mass of the Higgs boson (or any other undiscovered
remnant of electroweak symmetry breaking), the poorly known parameters
all involve quarks.
Other than top,\cite{CDF94} which decays too quickly for confinement
to play a role, the masses of the quarks are a bit better than wild
guesses.
The information on the Cabibbo-Kobayashi-Maskawa (CKM) quark-mixing
matrix is spotty, especially when one relaxes the assumption of
three-generation unitarity, as shown in Table~\ref{table:ckm}.
\begin{table}
\caption[table:ckm]{Ranges for CKM matrix elements $|V_{qr}|$ assuming
unitarity but {\em not\/} three generations.
Numerical values taken from the 1994 Review of Particle Properties.
In three generations $|V_{ud}|=s_{12}$, $|V_{cb}|=s_{23}$, and
$|V_{ub}|=s_{13}$, to excellent approximation.}\label{table:ckm}
\begin{center} \begin{tabular}{c@{\hspace{2.0em}}c@{\hspace{2.0em}}c}
\hline \hline
   parameter       & value or range & related lattice calculations \\
\hline
$|V_{ud}|$         &      0.974       &         \\
$|V_{us}|$         &   0.218--0.224   & $K\to \pi e\nu$ \\
$|V_{ub}|$         &   0.002--0.005   & $B\to \pi l\nu$ \\
$|V_{cd}|$         &   0.180--0.228   & $D\to \pi l\nu$  \\
$|V_{cs}|$         &   0.800--0.975   & $D\to  K  l\nu$ \\
$|V_{cb}|$         &   0.032--0.048   & $B\to D^* l\nu$ \\
$|V_{td}|$         &   0.0--0.13      & $f_B^2B_B$; $B_K$  \\
$|V_{ts}|$         &   0.0--0.56      & $f_{B_s}^2B_{B_s}$ \\
$|V_{tb}|$         &   0.0--0.9995    &         \\
\hline \hline
\end{tabular} \end{center}\end{table}
They are poorly determined simply because experiments measure
properties not of quarks, but of the hadrons inside which they
are confined.
Of course, everyone knows what to do: calculate with QCD, the part
of the Standard Model that describes the strong interactions.
But then, the strong coupling is known only at the 5\% level;
not bad, but nothing like the fine structure or Fermi constants.
Moreover, the binding of quarks into hadrons is nonperturbative---the
calculations cannot be done on the back of an envelope.

The most systematic technique for understanding nonperturbative QCD is
lattice gauge theory.
The lattice provides quantum field theory with a consistent and
mathematically well-defined ultraviolet regulator.
At fixed lattice spacing, the quantities of interest are straightforward
(combinations of) functional integrals.
These integrals can be approximated by a variety of techniques borrowed
from statistical mechanics.
Especially promising is a numerical technique, the Monte Carlo method
with importance sampling, which has become so pre-eminent that the
young and uninitiated probably haven't heard of any other.

Results from lattice-QCD Monte Carlo calculations have begun to
influence Table~\ref{table:sm}.
The world average for the SU(3) gauge coupling $\alpha_S$
includes results from lattice calculations of the quarkonium
spectrum,\cite{Kha92,Lep94,Kha95} and at the time of this conference an
even more precise result had appeared.\cite{Dav95}
The same calculations are also providing some of the best information
on the charm\cite{Kha94} and bottom\cite{Dav94} masses.
This is an auspicious beginning.
Over the next several years the lattice QCD calculations will mature.
They will help to determine the other unknowns---light quark masses
and the CKM matrix.
The third column of Tables~\ref{table:sm} and~\ref{table:ckm} lists
relevant quantities or processes, and the rest of this talk explains
how the program fits together.

Sect.~\ref{LGT} gives the non-expert some perspective on the conceptual
and numerical strengths and weaknesses of lattice QCD. 
Sect.~\ref{QCD} reviews 1) the status of the light hadron spectrum and
the propects for extracting $m_u$, $m_d$, and $m_s$; and 2) results for
the quarkonium spectrum, which yield $\alpha_S$, $m_b$, and $m_c$.
Sect.~\ref{EW} outlines lattice QCD calculations of electroweak,
hadronic matrix elements that are needed to pin down the unitarity
triangle of the CKM matrix.
There are, of course, many other interesting applications to electroweak
phenomenology; for more comprehensive reviews the reader can consult
some of the papers listed in the bibliography.\cite{Kro93,EWR94}

\section{Rudiments of Lattice Gauge Theory}\label{LGT}
In quantum field theory physical measurements are related to vacuum
expectation values $\langle \cO\rangle$.
Feynman's functional integral representation is
\begin{equation}\label{eq:path-integral}
\langle \cO\rangle = \lim_{L\to\infty} \lim_{a\to0} Z^{-1}
\int\prod_{x,\mu}dA_\mu(x)\prod_{x,a} d\psi_a(x)d\bar{\psi}_a(x)
\,\cO\,e^{-S(A_\mu,\psi,\bar{\psi})}.
\end{equation}
The formula is easier to understand if read from right to left.
$S$ is the action---in our case the action of QCD, so it depends on the
gluons $A_\mu$ and the quarks $\psi$ and anti-quarks $\bar{\psi}$.
$\cO$ depends on the physics under investigation; the most useful kinds
of $\cO$'s are given below.
The integration over all components and positions of the basic fields,
with weight $e^{iS}$ would reproduce the familiar Schr\"odinger or
Heisenberg formulations of quantum mechanics.
The more convergent weight $e^{-S}$ provides some benefits and imposes
some restrictions---see below.
$Z$ is the same integral without $\cO$ in it,
so that $\langle1\rangle=1$.

The limits are there to satisfy the mathematicians; without them the
integrals are not well defined.
These ``cutoffs'' also have a physical significance: we do not claim to
understand physics either at distances smaller than $a$ (the ultraviolet
cutoff), or at distances larger than $L$ (the infrared cutoff).
The limit $a\to0$ requires the renormalization group; it must be carried
out holding $L$ and physical, infrared scales fixed.
In particular, the integration variables $A_\mu(x)$, $\psi(x)$, and
$\bar{\psi}(x)$ really represent all degrees of freedom in a block of
size $a^4$.
The limit ``$a\to0$'' can be obtained not only literally, but also by
improving the action of the blocked fields.

These observations apply to any cutoff scheme for quantum field theory.
A nice introduction to the renormalization-group aspects is a
summer-school lecture by Lepage.\cite{Lep90}
In lattice gauge theory $a$ is nothing but the spacing between lattice
sites.
If there are $N$ on a side, $L=Na$.
For given $N$ one can compute the integrals numerically.
With the $10^7$--$10^{10}$-dimensional integrals that arise, the only
viable technique is a statistical one:
Monte Carlo with importance sampling.

To compute masses the observable $\cO=\Phi(t)\Phi^\dagger(0)$, where
$\Phi(t)$ is an operator at time $t$ with the flavor and
angular-momentum quantum numbers of the state of interest.
One can construct such operators using symmetry alone.
The radial quantum number would require a solution of the theory,
but that's what we're after.
The ``two-point function''
\begin{equation}\label{eq:two-point}
\langle \Phi(t)\Phi^\dagger(0) \rangle =
\sum_n \left|\langle0|\Phi|n\rangle\right|^2 e^{-m_nt},
\end{equation}
where the sum is over the radial quantum number.
The exponentials are a happy consequence of the weight $e^{-S}$
in eq.~(\ref{eq:path-integral}).
It is advantageous because at long times $t$ only the lowest-lying
state survives.
In a numerical calculation masses are obtained by fitting two-point
functions, once single-exponential behavior is verified.
Since $\Phi$ is largely arbitrary, some artistry enters: if
single-exponential behavior sets in sooner, the statistical quality of
the mass estimate is better.

To compute a matrix element of part of the electroweak Hamiltonian,
$\cH$, the observable $\cO=\Phi_\pi(t_\pi)\cH(t_h)\Phi_B^\dagger(0)$
for the transition from hadron ``$B$'' to hadron ``$\pi$.''
At long times $t_h$ and $t_\pi-t_h$ the ``three-point function''
\begin{equation}\label{eq:three-point}
\langle \Phi_\pi(t_\pi)\cH(t_h)\Phi_B^\dagger(0) \rangle \approx
\langle0|\Phi_\pi|\pi\rangle e^{-m_\pi(t_\pi-t_h)}
\langle\pi|\cH|B\rangle e^{-m_Bt_h} \langle B |\Phi_B^\dagger|0\rangle,
\end{equation}
plus excited-state contributions.
If, as in decays of hadrons to leptons, the hadronic final-state is the
vacuum, a two-point function will do:
\begin{equation}\label{eq:H-point}
\langle \cH(t)\Phi_B^\dagger(0) \rangle =\sum_n
\langle0|\cH|B_n\rangle e^{-m_nt}
\langle B_n|\Phi_B^\dagger|0\rangle.
\end{equation}
The desired matrix elements $\langle\pi|\cH|B\rangle$ and
$\langle0|\cH|B\rangle$ can be obtained from eq.~(\ref{eq:three-point})
and~(\ref{eq:H-point}), because the masses and $\Phi$-matrix elements
are obtained from eq.~(\ref{eq:two-point}).

To obtain good results from
eqs.~(\ref{eq:two-point})--(\ref{eq:H-point}), it is crucial to devise
nearly optimal operators in the two-point analysis.
Consumers of numerical results from lattice QCD should be wary of
results, still too prevalent in the literature, that are contaminated by
unwanted states.


In the numerical work that mostly concerns us here, the integrals are
computed at a sequence of fixed $a$'s and $L$'s.
One adopts a standard mass, say $m_\rho$, and defines
\begin{equation}\label{eq:units}
a=\frac{(am_\rho)^{\rm lQCD}}{m_\rho^{\rm expt}}
\end{equation}
to obtain the latice spacing in physical units, and other quantities are
predicted via
\begin{equation}
m_B=\frac{(am_B)^{\rm lQCD}}{a}.
\end{equation}
For continuum-limit, infinite-volume results this is the same as
extrapolating dimensionless ratios, e.g.\
\begin{equation}
\frac{m_B}{m_\rho}= \lim_{L\to\infty} \lim_{a\to0}
\frac{am_B(a,L)}{am_\rho(a,L)}.
\end{equation}
There is theoretical guidance for both limits.
According to general properties of massive quantum field theories in
finite boxes,\cite{Lue86} the infinite-volume limit is rapid for
$m_\pi L\gg1$, exponential for masses.
In non-Abelian gauge theories the renormalization-group $a\to0$ limit is
controlled by asymptotic freedom.

The main strength of lattice QCD is that it {\em is\/} QCD.
It has $1+n_f$ adjustable parameters, corresponding to the gauge
coupling and the quark masses.
From the renormalization group, the adjustment of the gauge
coupling is equivalent to setting the lattice spacing in physical
units, eq.~(\ref{eq:units}).
Once the parameters are determined by $1+n_f$ experimental inputs,
QCD should predict all other strong-interaction phenomena.
There is no need to introduce condensates, as in ITEP sum rules,
or non-renormalizable couplings, as in chiral perturbation theory or
heavy-quark effective theory.
If theory and experiment disagree, it is a signal of new physics.

There are some disadvantages.
A practical, though not conceptual, problem is that large-scale
computational work is more labor-intensive than traditional theoretical
physics.
Careful work is needed to estimate the uncertainties reliably.
The improvements in computer power and algorithms of recent years have
helped practitioners understand their uncertainties better and better.
As the consumers of their results become commensurately sophisticated,
this trend will continue.
After all, in the context of Table~\ref{table:sm}, meaningful error bars
are just as important as the central value.

\subsection{The quenched approximation}\label{sect:quenched}
The biggest disadvantage of most of the numerical results mentioned in
this talk is something called the ``quenched'' approximation.
A meson consists of a valence quark and anti-quark exchanging any number
of gluons.
The gluons can turn into virtual quark loops and back again.
The latter process costs a factor of 100-1000 in computer time, so many
Monte Carlo programs just omit the virtual quark loops.
To accentuate the positive---gluons and valence quarks are treated
better than in non-QCD models of hadrons---the omission is sometimes
called the {\em valence\/} approximation.
To admit the negative, it is less often called the {\em loopless\/}
approximation.
But most often lattice mavens borrow a jargon from condensed-matter
physics and call it the {\em quenched\/} approximation.

If quenched QCD makes any sense, it is as a kind of model or effective
theory.
The parameters of quenched QCD can be tuned to reproduce physics at one
scale.
But the $\beta$ function of quenched and genuine QCD differ, as one sees
in perturbation theory, so one cannot expect agreement at all
scales.
As with any model, only in special cases can one argue that these
effects are correctable or negligible; these cases will be highlighted
in the rest of the talk.

\section{From Hadron Spectra to the QCD Parameters}\label{QCD}
\subsection{Light hadrons and light-quark masses}
Over the past few years a group at IBM has carried out a systematic
calculation of the light-hadron spectrum using the dedicated
supercomputer GF11.\cite{But93}
They have numerical data for 5 different combinations of $(a,L)$.
At $L\approx2.3$~fm there are three lattice spacings varying by a factor
of $\sim2.5$.
At the coarsest lattice spacing ($a^{-1}\approx1.4$~GeV) there are three
volumes, up to almost 2.5~fm.
A variety of quark masses are used, and the physical strange quark is
reached by interpolation, whereas the light (up and down) through
extrapolation.
The mass dependence is assumed linear, as expected from weakly broken
chiral symmetry, and the data substantiate the assumption.

\begin{figure}
\epsfxsize=\textwidth \epsfbox{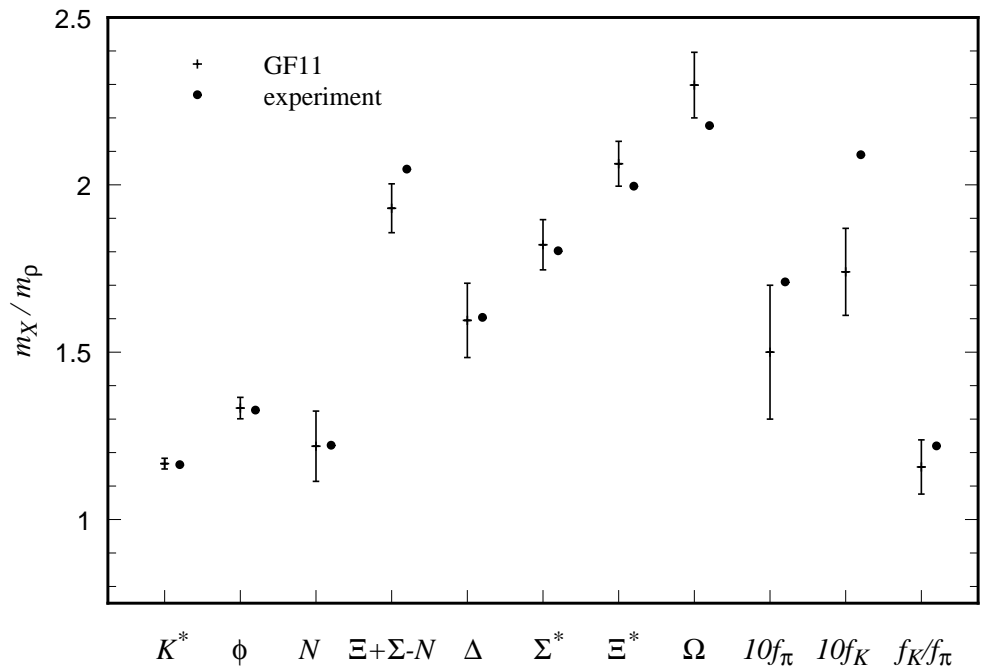}
\caption[fig:spectrum]{The spectrum and decay constants of the light
hadrons.  Error bars are from lattice calculations in the quenched
approximation,\cite{But93,But94} and $\bullet$ denotes
experiment.}\label{fig:spectrum}
\end{figure}

The units (i.e.\ lattice spacing) has been fixed with $m_\rho$ and the
quark masses with $m_\pi$ and $m_K$.
The final results, after extrapolation to the continuum limit and
infinite volume, are shown in Fig.~\ref{fig:spectrum} for two vector
mesons and six baryons.
(The quark-mass interpolation could reach only the combination
$m_\Xi+m_\Sigma-m_N$.)
Despite the quenched approximation the agreement with experiment is
spectacular.

Fig.~\ref{fig:spectrum} also includes results from the same
investigation for decay constants.
The agreement of $f_\pi/m_\rho$ and $f_K/m_\rho$ is not as good as for
the masses.
Because of the quenched approximation, this is not entirely unexpected.
Recall the argument concerning distance scales and effective theories in
sect.~\ref{sect:quenched}.
The binding mechanism responsible for the masses encompasses distances
out to the typical hadronic radius.
The decay constant, on the other hand, is proportional to the
wavefunction at the origin and thus is more sensitive to shorter
distance scales.
One sees better agreement when forming the ratio $f_K/f_\pi$,
which---recall eq.~(\ref{eq:units}) and subsequent discussion---is
like retuning to the shorter distance.

One would like to use the hadron masses to extract the quark masses.
Because of confinement, the quark mass more like a renormalized coupling
than the classical concept of mass.
Calculations like the one described above yield immediately the
bare mass of the lattice theory.
More useful to others would be the $\MSbar$ scheme of dimensional
regularization.
A one-loop perturbative calculation can be used to convert from one
scheme to another.\cite{GHS84,Mor93,Kro94,Kha94}

For the light quarks it is convenient to discuss the combinations
$\hat{m}=\half(m_d+m_u)$, $\Delta m^2_{du}=m^2_d-m^2_u$, and $m_s$.
Ratios of the light-quark masses are currently best estimated using
chiral perturbation theory.\cite{Gas82}
To set the overall scale requires a dynamical calculation in QCD.
In lattice QCD, $\hat{m}$ and $m_s$ can be extracted from the variation
in the square of the pseudoscalar mass between $m_\pi^2$ and $m_K^2$.
The most difficult quark-mass combination is $\Delta m^2_{du}$, which
causes the isospin-violating end of the splittings in hadron multiplets.
Since chiral perturbation theory provides a formula for
$\Delta m^2_{du}/m_s^2$ with only second-order corrections, it is
likely that the best determination of $\Delta m^2_{du}$ will come from
combining the formula with a lattice QCD result for $m_s$.

Using the compliation of quenched and unquenched results of
Ukawa,\cite{Ukawa} Mack\-en\-zie\cite{Mac94} has estimated
$\hat{m}_\MSbar(1~{\rm GeV})\sim2.3$~MeV and
$m_{s,\MSbar}(1~{\rm GeV})\sim65$~MeV.
The symbol $\sim$ stresses the lack of error bar.
This is outside the ranges of 3.5--11.5~MeV and 100--300~MeV
indicated in Table~\ref{table:sm}.
A more recent analysis of the strange quark finds
$m_{\MSbar,s}(2~{\rm GeV})=127\pm18$~MeV,\cite{All94} in the lower
part of the range in Table~\ref{table:sm}.
None of these results should be taken seriously until a more complete
error analysis exists, but it is intriguing that the conventional
estimates might be too high.

\subsection{Quarkonia, $\alpha_S$, and heavy-quark masses}
Quarkonia are bound states of a heavy quark and heavy anti-quark.
Three families of states exist, charmonium ($\eta_c$, $J/\psi$, etc),
bottomonium ($\eta_b$, $\Upsilon$, etc), and the as yet unobserved $B_c$
($b\bar{c}$ and $\bar{b}c$ bound states).
Compared to light hadrons, these systems are simple.
The quarks are nonrelativistic, and potential models give an excellent
empirical description.
But a fundamental treatment of these systems requires nonperturbative
QCD, i.e.\ lattice QCD.%
\footnote{The utility of quarkonia for testing the methodology of
lattice gauge theory and the theory of QCD has been stressed over and
over by Peter Lepage.}
Potential models can be exploited, however, to estimate lattice
artifacts, and in the quenched approximation they can be used to make
corrections.
Many states have been observed in the lab, providing cross-checks of
the methodology of uncertainty estimation.

Once the checks are satisfactory, one can use the spectra to determine
$\alpha_S$, $m_c$, and $m_b$.
One can also have some confidence in further applications, such as the
phenomenology of $D$ and $B$ mesons discussed in sect.~\ref{EW}.

For charm, and especially for bottom, the quark mass is close to the
ultraviolet cutoff, $1/a$ or $\pi/a$, of present-day numerical
calculations.
Originally lattice gauge theory was formulated assuming $m_qa\ll1$,
so quarks $m_qa\sim1$ require some reassessment.
There are four ways to react.
The patient, stolid way is to wait ten years, until computers are
powerful enough to reach a cutoff of 20~GeV---not very inspiring.
The naive way is to extrapolate from smaller masses, assuming the
$m_qa\ll1$ interpretation of the lattice theory is adequate; history
shows that naive extrapolations can lead to naive and, thus,
unacceptable error estimates.
The insightful way is to formulate an effective theory for heavy quarks
with a lattice cutoff;\cite{Lep87,Eic87} this is the computationally
most efficient approach, and when the effectiveness of the heavy-quark
expansion is {\em a~priori\/} clear, it is the method of choice.
The compulsive way to examine a wide class of lattice theories without
assuming either $m_qa\ll1$ or $m_q\gg(\LQCD,a^{-1})$; by imposing
physical normalization conditions on masses and matrix elements,
it is possible to interpret the correlation functions at {\em any\/}
value of $m_qa$.\cite{KKM9?}
The underlying reason is that the lattice theory is completely
compatible with the heavy-quark limit, so the mass-dependent
interpretation connects smoothly onto both the insightful method for
$m_qa\gg1$ and the standard method for $m_qa\ll1$.

\begin{figure}
\epsfxsize=\textwidth \epsfbox{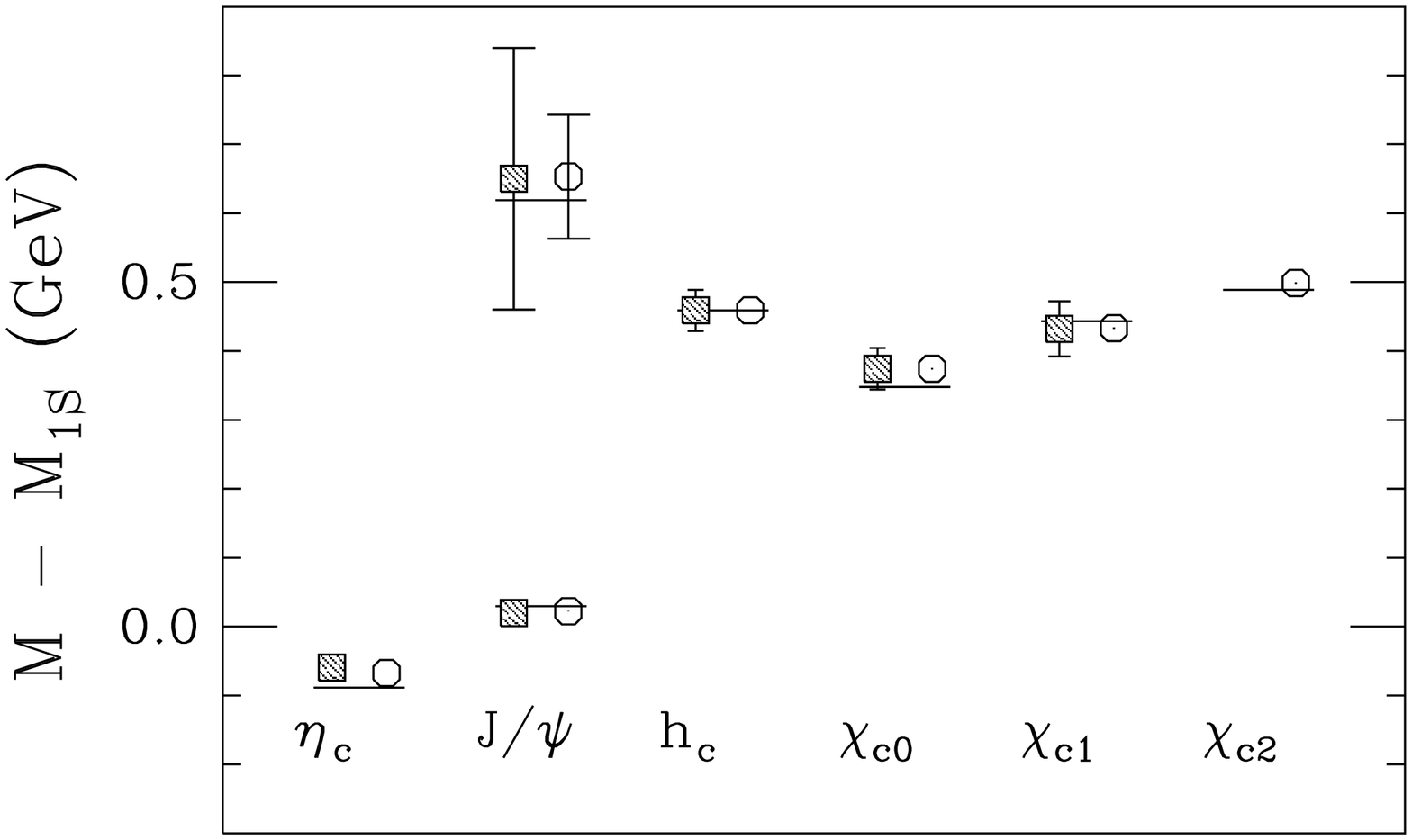}
\caption[fig:cbarc]{A comparison of the charmonium spectrum as
calculated in lattice QCD, using two different methods.
\refcite{NRQ94}: $\circ$, \refcite{Kha93}: $\Box$.
From \refcite{Kha95}.}\label{fig:cbarc}
\end{figure}

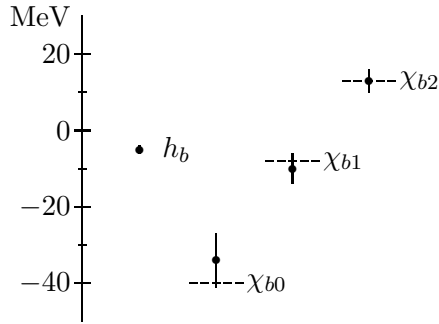
\begin{figure}
\begin{center}
\setlength{\unitlength}{0.02in}
\begin{picture}(110,140)(0,-50)
\put(15,-50){\line(0,1){80}}
\multiput(13,-40)(0,20){4}{\line(1,0){4}}
\multiput(14,-40)(0,10){7}{\line(1,0){2}}
\put(12,-40){\makebox(0,0)[r]{$-40$}}
\put(12,-20){\makebox(0,0)[r]{$-20$}}
\put(12,0){\makebox(0,0)[r]{$0$}}
\put(12,20){\makebox(0,0)[r]{$20$}}
\put(12,30){\makebox(0,0)[r]{MeV}}

\multiput(43,-40)(3,0){5}{\line(1,0){2}}
\put(58,-40){\makebox(0,0)[l]{$\chi_{b0}$}}
\put(50,-34){\circle*{2}}
\put(50,-34){\line(0,1){7}}
\put(50,-34){\line(0,-1){7}}

\put(36,-5){\makebox(0,0)[l]{$h_b$}}
\put(30,-5){\circle*{2}}
\put(30,-5){\line(0,1){1}}
\put(30,-5){\line(0,-1){1}}

\multiput(63,-8)(3,0){5}{\line(1,0){2}}
\put(78,-8){\makebox(0,0)[l]{$\chi_{b1}$}}
\put(70,-10){\circle*{2}}
\put(70,-10){\line(0,1){4}}
\put(70,-10){\line(0,-1){4}}

\multiput(83,13)(3,0){5}{\line(1,0){2}}
\put(98,13){\makebox(0,0)[l]{$\chi_{b2}$}}
\put(90,13){\circle*{2}}
\put(90,13){\line(0,1){3}}
\put(90,13){\line(0,-1){3}}
\end{picture}
\end{center}
\caption[fig:hyperfine]{Lattice QCD results for the spin-dependent
splittings of the lowest-lying P states in bottomonium.
The dashed lines are the experimental values, where available.
Energies are measured relative to the spin average of the $\chi$ states.
From \refcite{NRQ94}.}\label{fig:hyperfine}
\end{figure}

Fig.~\ref{fig:cbarc} shows the charmonium spectrum, on a scale
appropriate to the spin-averaged spectrum.
Light quark loops are quenched in these calculations.\cite{NRQ94,Kha93}
The agreement with experimental measurements is impressive, but
Fig.~\ref{fig:cbarc} barely displays the attainable precision.
Fig.~\ref{fig:hyperfine} shows the fine and hyperfine structure of the P
states, now for bottomonium.\cite{NRQ94}
(The $^1$P$_1$ state $h_b$ has not been observed in the lab; the
$h_c$ has been seen.)
The authors of \refcite{NRQ94} also have results with the virtual quark
loops from two light quarks, i.e.\ up and down are no longer quenched,
but strange still is.
The agreement is comparable.\cite{Slo94}

To obtain these results only two parameters have been adjusted.
The standard mass in eq.~(\ref{eq:units}) is $\mps$, the spin-averaged
splitting of the 1P and 1S states, which is insensitive to the quark
mass.
By the renormalization group, this is equivalent to eliminating the bare
gauge coupling, or to determining $\LQCD$.
The bare quark mass is adjusted to obtain the spin average of the 1S
states that is measured in the lab.
Otherwise figs.~\ref{fig:cbarc} and~\ref{fig:hyperfine} represent
predictions of quenched QCD.

The success of these calculations permits one to extract the basic
parameters, $\alpha_S$ and $m_q$.
There are four steps:
\begin{enumerate}
\item\label{compute}
Compute the charm- and bottomonium spectra with $n_{f,\rm MC}=0,2$ or 3
flavors of virtual quark loops.
($n_{f,\rm MC}=0$ corresponds to the quenched approximation;
 $n_{f,\rm MC}=2$ quenches just the strange quark;
 $n_{f,\rm MC}=3$ would be the real world.)
\item\label{convert}
With perturbation theory, convert the bare lattice coupling
$\alpha_0^{(n_{f,\rm MC})}$ to the quark-potential ($V$) or $\MSbar$
scheme; convert the bare lattice mass $(m_0a)^{(n_{f,\rm MC})}$ to the
pole or $\MSbar$ scheme.
The natural scale for this conversion is near
(but not quite\cite{Lep93}) $\pi/a$.
\item\label{correct}
Unless $n_{f,\rm MC}=3$, correct for the quenched approximation.
\item\label{amps}
Eliminate $a$ from $\alpha_\MSbar(\pi/a)$ and $am_\MSbar(\pi/a)$ using
\begin{equation}
a=\frac{a\mps}{460~{\rm MeV}},
\end{equation}
where the numerator is the 1P--1S splitting in lattice units.
\end{enumerate}
Steps \ref{compute} and \ref{amps} are explained above.
Step~\ref{convert} requires one-loop perturbation theory, suitably
optimized.\cite{Lep93}
Step~\ref{correct} is crucial, because without it the results have no
business in Table~\ref{table:sm}.

Consider first $\alpha_S$, and recall the idea of treating the quenched
approximation as an effective theory.
One sees that the couplings are implicitly matched at some scale $q_Q$
characteristic of quarkonia.
So the matching hypothesis, supported by figs.~\ref{fig:cbarc}
and~\ref{fig:hyperfine}, asserts
\begin{equation}
\alpha_S^{(n_{f,\rm MC})}(q_Q)=\alpha_S^{(3)}(q_Q).
\end{equation}
Potential models tell us that $200<q_c<800$~MeV and $200<q_b<1400$~MeV.
Step 3 yields $\alpha_S^{(n_{f,\rm MC})}(\pi/a)$, so one can use the
two-loop perturbative renormalization group to run from $\pi/a$ to
$q_Q$.
The perturbative running is an overestimate if $q_Q$ is taken at the
lower end of these ranges.%
\footnote{For light hadrons, $q_{\rm light}\sim\LQCD$, so there would
be no perturbative control whatsoever.}
This argument was used for the original lattice determinations of the
strong coupling,\cite{Kha92,Lep94} and its reliability was confirmed in
$n_{f,\rm MC}=2$ calculations.\cite{Aok95}

Currently the most accurate result is from \refcite{Dav95},
\begin{equation}\label{eq:alpha_V}
\alpha_V^{(3)}(8.2~{\rm GeV})=0.196\pm0.003,
\end{equation}
based on $n_{f,\rm MC}=0$ and $n_{f,\rm MC}=2$ results, with an
extrapolation in $n_f$.
The $V$ scheme is preferred for the matching argument, not only for
physical reasons, but also because of its empirical scaling
behavior.\cite{Lep93}
The scaling behavior implies that one can run with the two-loop
renormalization group to high scales and convert to other schemes.
For comparison to other determinations,
eq.~(\ref{eq:alpha_V}) corresponds to
\begin{equation}\label{eq:alpha_MSbar}
\alpha_\MSbar(M_Z)=0.115\pm0.002.
\end{equation}
The quoted uncertainty is smaller than that reported from any other
method.
The largest contributor is the quenched correction; the second largest
is the perturbative conversion $0\to V\to\MSbar$.

To determine the quark mass the one applies the same
renormalization-group argument.
But, quark masses don't run below threshold!\cite{Dav94}
Hence, for heavy quarks%
\footnote{For light quarks the threshold is deep in brown muck, and all
bets are off.}
$m_Q^{(n_{f,\rm MC})}(m_Q)=m_Q^{(n_{f,\rm MC})}(q_Q)=%
 m_Q^{(3)}(q_Q)=m_Q^{(3)}(m_Q)$.
The only corrections are perturbative, from lattice conventions to
$\MSbar$ or physical conventions.
Using the convention of the perturbative ``pole mass''
\begin{equation}\label{eq:mq}
\begin{array}{r@{\,=\,}l@{\;\rm MeV\;}l}
m_c & 1.5\pm0.2 & \mbox{(\refcite{Kha94}, preliminary)},\\[1.0em]
m_b & 5.0\pm0.2 & \mbox{(\refcite{Dav94})}.
\end{array}
\end{equation}
At the nonperturbative level confinement wipes out the pole, so the name
``pole mass'' should not be taken too literally.
The perturbative pole mass is like a running mass, except that it
has a fixed scale built into the definition.
It is useful, because it is thought\cite{Dav95} to correspond to the
mass of phenomenological models that do not probe energies less than
$\LQCD$, such as potential models.
In other contexts---such as the study of Yukawa couplings in unification
scenarios---the $\MSbar$ convention may be more appropriate.
Eq.~(\ref{eq:mq}b) corresponds to $m_{b,\MSbar}(m_b)=4.0\pm0.1$~GeV.

%

\section{From Matrix Elements to the CKM Matrix}\label{EW}
Electroweak decays of flavored hadrons follow the schematic formula
\begin{equation}\label{eq:factors}
\left( \begin{array}{c} {\rm experimental} \\
                        {\rm measurement}  \end{array} \right) =
\left[ \begin{array}{c} {\rm known} \\ {\rm factors} \end{array} \right]
\left( \begin{array}{c} {\rm  QCD}  \\ {\rm factor}  \end{array} \right)
\left( \begin{array}{c} {\rm  CKM}  \\ {\rm factor}  \end{array} \right)
\end{equation}
North American, Japanese, and European taxpayers provide us with lots of
money for the relevant experiments, because they want to know the CKM
factors.
But unless we calculate the inherently nonperturbative QCD factor,
they will be sorely disappointed.

It is convenient to start with the assumption of three-generation
unitarity.
Then
\begin{equation}\label{eq:triangle}
V_{ud}V_{ub}^* + V_{cd}V_{cb}^* + V_{td}V_{tb}^* = 0,
\end{equation}
an equation that prescribes a triangle in the complex plane.
Dividing by $V_{cd}V_{cb}^*$ and writing
$V_{ud}V_{ub}^*/V_{cd}V_{cb}^*=\bar{\rho}+i\bar{\eta}$, one sees that
unitarity predicts
\begin{equation}
\frac{V_{td}V_{tb}^*}{V_{cd}V_{cb}^*} = 1-\bar{\rho}-i\bar{\eta}.
\end{equation}
The notation\cite{BLO94} $(\bar{\rho},\bar{\eta})$ is to distinguish
these parameters from the standard Wolfenstein parameters
$(\rho,\eta)=(\bar{\rho},\bar{\eta})/|V_{ud}|$.
The standard CKM phase in Table~\ref{table:sm} is
$\delta=\tan^{-1}(\eta/\rho)=\tan^{-1}(\bar{\eta}/\bar{\rho})$.

One would like to  determine $(\bar{\rho},\bar{\eta})$ through as many
physical processes as possible.
For example, the strength of $CP$ violation in $b$-flavored hadrons is
related to the angles of the triangle, hence the high interest in $B$
factories.
If all experiments agree, the test verifies the CKM explanation of $CP$
violation; if not, the discrepency would have to be explained by physics
beyond the Standard Model.
Meanwhile, lattice QCD is useful for measuring the sides of the
unitarity triangle.
Depending on the shape of the triangle, the precision may be good enough
to predict the angles before the $B$ factories have been built.

Let us consider the CKM matrix elements in eq.~(\ref{eq:triangle}).
Assuming three-generation unitarity, $1-|V_{tb}|$ is too small to
worry about, and $|V_{cd}|-|V_{us}|$ is also very small.
In principle, however, $|V_{us}|$ and $|V_{cd}|$ can be determined from
lattice QCD and measurements of the semi-leptonic decays $K\to\pi e\nu$
and $D\to\pi l\nu$, respectively.
The technique is the same as for $|V_{ub}|$ from $B\to\pi l\nu$,
discussed below.
It is unlikely that lattice QCD can, or will need to, improve on
$|V_{ud}|=0.9744\pm0.0010$ during the period relevant to this
discussion.

The most poorly known elements of eq.~(\ref{eq:triangle}) are
$|V_{cb}|$, $|V_{ub}|$, and $|V_{td}|$.
In principle, the first two can be determined from leptonic decays
$B_q\to\tau\nu$, $q=u,\,b$, but the experimental prospects are bleak.
The semi-leptonic decay is more promising.
Near $q^2_{\rm max}=(m_B-m_{D^*})^2$ the differential decay rate for
$B\to D^*l\nu$ is
\begin{equation}\label{eq:semi-vector}
\frac{d\Gamma}{dq^2}= \left[\frac{G_F^2q^2}{64\pi^3m_B}
\Big((q^2_{\rm max}-q^2)(4m_{D^*}m_B + q^2_{\rm max}-q^2)\Big)^{1/2}
\right] |A_1(q^2)|^2|V_{cb}|^2,
\end{equation}
where $q^2$ is the invariant mass-squared of the lepton system.
One must carry out a nonperturbative QCD calculation to obtain
the form factor $A_1(q^2)$.
By heavy-quark symmetry, however, $A_1(q^2_{\rm max})$ obeys a
normalization condition, up to $1/m_{D^*}^2$ corrections\cite{Luk90}
(estimated to be small) and known radiative corrections.
Other form factors, which are phase-space suppressed near
$q^2_{\rm max}$, are also related by heavy-quark symmetry to $A_1(q^2)$.
Hence, eq.~(\ref{eq:semi-vector}) provides an essentially
model-independent\cite{Neu91} way to determine $|V_{cb}|$.

\begin{figure}
\epsfxsize=\textwidth \epsfbox{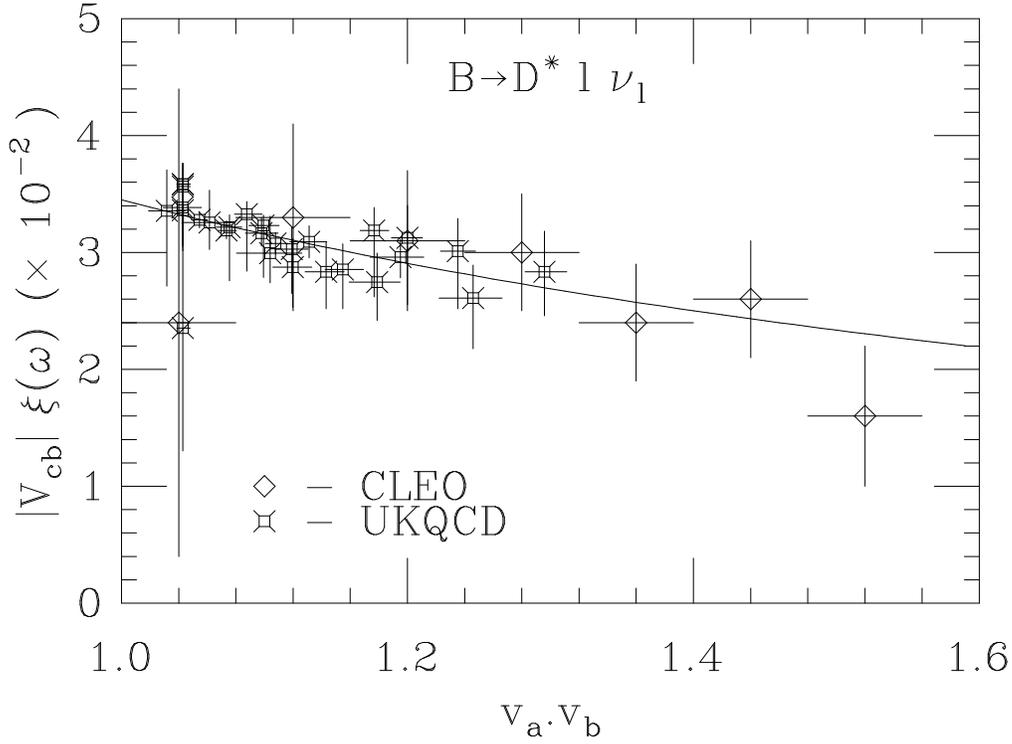}
\caption[fig:simone]{The Isgur-Wise function $\xi(\omega)$
(essentially the form factor $A_1$ of the text) from lattice QCD
and CLEO.  The kinematic variable
$\omega=v_a\cdot v_b=1-(q^2_{\rm max}-q^2)/2m_Bm_{D^*}$.
From \refcite{Sim94}.}\label{fig:simone}
\end{figure}

The difficulty with the model-independent analysis is that the decay
rate vanishes at $q^2_{\rm max}$.
To aid experimentalists' extrapolation to that point, several
groups\cite{Ber93,Boo94,Man94}
have used quenched lattice QCD to compute the slope of $A_1$.
A typical analysis is to fit the slope to lattice-QCD numerical data,
and then fit the normalization to CLEO's experimental data, as shown in
Fig.~\ref{fig:simone}.
For example, Simone of the UKQCD Collaboration finds\cite{Sim94}
\begin{equation}\label{eq:Vcb}
|V_{cb}|=0.034^{+3+2}_{-2-2}\sqrt{\frac{\tau_B}{1.49~\rm ps}}.
\end{equation}
The first error is experimental; the second is from the lattice-QCD
slope.
Unfortunately, it is not clear how to correct for the quenched
approximation, and the associated uncertainty has not been estimated.
Moreover, consistency checks of varying lattice spacing, volume, etc,
are still in progress.
Nevertheless, the overall consistency with experiment, shown in
Fig.~\ref{fig:simone}, is encouraging.

$|V_{ub}|$ can be obtained from the semi-leptonic decays
$B\to\rho l\nu$ and $B\to\pi l\nu$.
Expanding in $q^2$ near $q^2_{\rm max}=(m_B-m_\pi)^2$, the differential
decay rate for $B\to\pi l\nu$ reads
\begin{equation}\label{eq:semi-pseudoscalar}
\frac{d\Gamma}{dq^2} =
\left[\frac{G_F^2(q^2_{\rm max}-q^2)^{3/2}}{24\pi^3}
\left(\frac{m_\pi}{m_B}\right)^{3/2}\right]
|f_+(q^2_{\rm max})|^2|V_{ub}|^2
\left(1+{\rm O}(q^2_{\rm max}-q^2)\right),
\end{equation}
where $f_+(q^2)$ is the form factor that must be calculated in lattice
QCD.
Now, however, heavy-quark symmetry does not restrict
$f_+(q^2_{\rm max})$, so a calculation is needed to make any progress.
These calculations are underway at Fermilab, and, presumably, many
other places.

The third row of the CKM matrix can be probed via the box diagrams
responsible for neutral meson mixing.
The $CP=+$ admixture of the $K_L$ is parameterized by
$|\varepsilon_K|=2.26\times10^{-3}$.
The Standard Model predicts
\begin{equation}\label{eq:ek}
\begin{array}{l}
|\varepsilon_K|=
\left[\dfrac{\sqrt{2}G_F^2m_W^2}{16\pi^2m_K\Delta m_K}\right]
\mbox{\small$\frac{8}{3}$}m_K^2f_K^2\hat{B}_K
|V_{ud}V_{us}|^2|V_{cb}|^2 \times\\[1.0em] \hspace{6em}
\bar{\eta}\Big( |V_{cb}|^2(1-\bar{\rho}) y_t\eta_2f_2(y_t) +
y_c(\eta_3f_3(y_t)-\eta_1) \Big),
\end{array}
\end{equation}
where $y_q=m_q^2/m_W^2$.
This formula assumes three-generation unitarity and neglects the
deviation of $|V_{cs}|$ and $|V_{tb}|$ from unity.
The $\eta_i$ and $f_i$ multiplying the CKM factors arise from box
diagrams and their QCD corrections.\cite{Ina81}
The nonperturbative QCD factor is
$\mbox{\small$\frac{8}{3}$}m_K^2f_K^2B_K$, which is the $K$--$\bar{K}$
transition matrix element of a $\Delta S=2$ operator.
The best result for $B_K$ is\cite{Sha94,Ish93}
\begin{equation}\label{eq:B_K}
\begin{array}{r@{\,\pm\,}l}
B_K({\rm NDR,~2~GeV})=0.616 & 0.020\pm0.014\pm0.009\pm0.004\pm0.002
       \\[0.7em]            & ({\rm few~\%}) \pm3\%.
\end{array}
\end{equation}
The many error bars are exihibited to show how mature the uncertainty
analysis has become.
The first is statistical and the others are systematic.
The ``few~\%'' are for the quenched approximation.
This estimate comes from repeating some of the numerical computations
for full QCD,\cite{Kil92} though not enough to obtain the
other error bars, and from an analysis of chiral
logarithms.\cite{Sha90}
The latter study is reassuring only for degenerate quarks, so the
calculations are done with both quarks at $\half m_s$.
The 3\% uncertainty is an estimate of ${\rm O}(m_s-m_d)$ contributions.
Combining the errors and converting to the renormalization-group
invariant that appears in eq.~(\ref{eq:ek}), one finds\cite{Sha94}
\begin{equation}\label{eq:hatB_K}
\begin{array}{r@{\,=\,}l}
\hat{B}_K & (\alpha_\MSbar(\mu))^{-6/25}B_K(\rm NDR,~\mu) \\[0.7em]
          & 0.825\pm0.027(\,{\rm stat.}) \pm0.023(\,{\rm syst.})
\pm({\rm few~\%}) \pm3\%.
\end{array}
\end{equation}
This result places a high standard on calculations of $\hat{B}_K$,
whether by lattice QCD or any other method.
Would-be competitors must not only reach 10\% uncertainties, they must
do so with an error analysis as thorough and forthright as
\refcite{Sha94}.

Mixing in the $B^0$--$\bar{B}^0$ system is also sensitive to $V_{td}$.
In the Standard Model the mass splitting is given by
\begin{equation}\label{eq:xd}
x_d=\frac{\Delta m_{B_d}}{\Gamma_{B_d}}=
\left[\frac{G_F^2m_t^2\tau_{B_d}}{16\pi^2m_{B_d}}
\eta_Bf_2(y_t) \right]
\mbox{\small$\frac{8}{3}$}m_{B_d}^2f_{B_d}^2\hat{B}^{ }_{B_d}
|V_{td}^*V_{tb}|^2,
\end{equation}
The same formula holds for the $B_s$--$\bar{B}_s$ system, but with the
$d$ quark replaced by $s$
(i.e.\ $B_d\mapsto B_s$, $V_{td}\mapsto V_{ts}$.)
The nonperturbative QCD factor is
$\mbox{\small$\frac{8}{3}$}m_{B_q}^2f_{B_q}^2\hat{B}_{B_q}$,
which is the ${B_q}$--$\bar{B}_q$ transition matrix element of a
$\Delta B=2$ operator.
The calculation of the decay constant $f_B$ has received a great deal of
attention over the last several years,\cite{Som95} but the matrix
element needed here,
$\mbox{\small$\frac{8}{3}$}m_B^2f_B^2\hat{B}_B$,
has been mostly neglected.
(There are some older, exploratory papers.\cite{Ber88})

\begin{figure}
\epsfxsize=\textwidth \epsfbox{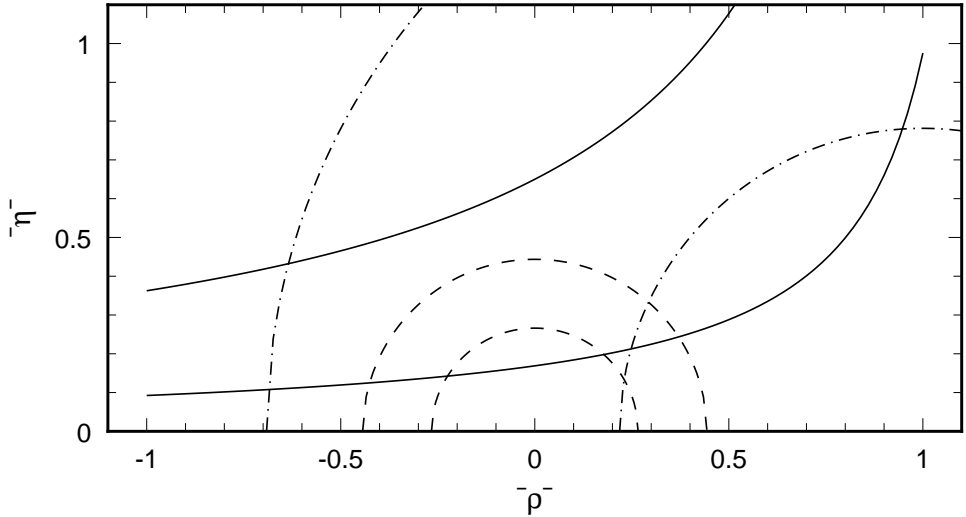}
\caption[fig:vtd]{Constraints on $(\bar{\rho},\bar{\eta})$ from
$|\varepsilon_K|$ (solid hyperbolae),
$|V_{ub}/V_{cb}|$ (dashed circles with origin (0,0)),
and $x_d/x_s$ (dash-dotted circles with origin (1,0)), and contemporary
uncertainties.}\label{fig:vtd}
\end{figure}

It is interesting to see how the lattice results influence the unitarity
triangle.
Fig.~\ref{fig:vtd} shows constraints from eqs.~(\ref{eq:ek}),
$|V_{ub}/V_{cb}|$, and $x_d/x_s$, taking for the masses
\begin{equation}
\begin{array}{r@{\,=\,}l}
m_{c,\MSbar}      & 1.3\pm 0.2, \\
m_{t,\MSbar}      & 175\pm 15,
\end{array}
\end{equation}
for the hadronic matrix elements
\begin{equation}
\begin{array}{r@{\,=\,}l}
\hat{B}_K         & 0.825\pm0.050, \\
|f_{B_d}/f_{B_s}| & 0.90\pm0.05, \\
|B_{B_d}/B_{B_s}| & 1.0\pm0.2,
\end{array}
\end{equation}
for ``experimental''%
\footnote{Nonperturbative QCD is needed to extract these results!}
CKM results
\begin{equation}
\begin{array}{r@{\,=\,}l}
|V_{cb}|          & 0.040\pm0.005, \\
|V_{ub}/V_{cb}|   & 0.08\pm0.02,
\end{array}
\end{equation}
and for neutral $B$ mixing measurements
\begin{equation}
\begin{array}{r@{\,=\,}l}
x_d           & 0.72\pm0.08, \\
x_s           & 15\pm5.
\end{array}
\end{equation}
Other inputs are as in \refcite{BLO94}.
I've made two wild guesses: $|B_{B_d}/B_{B_s}|$ and $x_s$.
But note that I take the uncertainty estimate in $\hat{B}_K$ seriously;
doubling it would not make much difference, in view of the uncertainties
in $m_t$ and $|V_{cb}|$.
Alas, these and the other uncertainties are too large to make
Fig.~\ref{fig:vtd} interesting.

\begin{figure}
\epsfxsize=\textwidth \epsfbox{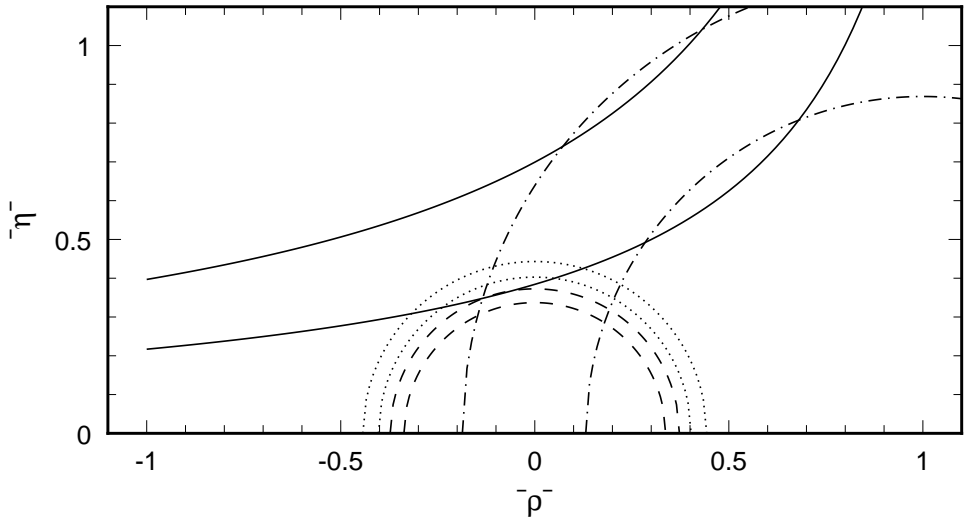}
\caption[fig:vtd5]{Constraints on $(\bar{\rho},\bar{\eta})$ from
$|\varepsilon_K|$ (solid hyperbolae),
low  $|V_{ub}/V_{cb}|$ (dashed circles with origin (0,0)) or
high $|V_{ub}/V_{cb}|$ (dotted circles with origin (0,0)),
and $x_d/x_s$ (dash-dotted circles with origin (1,0)), and improved
(5--10\%) uncertainties.}\label{fig:vtd5}
\end{figure}

What if lattice QCD calculation and the experiments improve?
Consider for the masses
\begin{equation}
\begin{array}{r@{\,=\,}l}
m_{c,\MSbar}      & 1.3\pm 0.1, \\
m_{t,\MSbar}      & 175\pm 5,
\end{array}
\end{equation}
for the hadronic matrix elements
\begin{equation}
\begin{array}{r@{\,=\,}l}
\hat{B}_K         & 0.825\pm0.027, \\
|f_{B_d}/f_{B_s}| & 0.90\pm0.02, \\
|B_{B_d}/B_{B_s}| & 1.0\pm0.1,
\end{array}
\end{equation}
in particular eliminating almost all the statistical error in
$\hat{B}_K$; for ``experimental'' CKM results
\begin{equation}
\begin{array}{r@{\,=\,}l}
|V_{cb}|          & 0.035\pm0.002, \\
|V_{ub}/V_{cb}|   & 0.080\pm0.004~~\mbox{``low,''}  \\
                  & 0.091\pm0.004~~\mbox{``high,''} \\
\end{array}
\end{equation}
and for neutral $B$ mixing measurements
\begin{equation}
\begin{array}{r@{\,=\,}l}
x_d           & 0.72\pm0.04,\\
x_s           & 18\pm2.
\end{array}
\end{equation}
Fig.~\ref{fig:vtd5} shows how this 5--10\% level of precision improves
the limits on $(\bar{\rho},\bar{\eta})$.
The wildest guess remains $x_s$, so ignore the dashed-dotted
curves momentarily.
The region allowed by the hyperbolic band from $\varepsilon_K$ and the
circular band from $|V_{ub}/V_{cb}|$ shrinks if $|V_{ub}|$ is too small.
The tension between these two constraints is partly a consequence of the
low value of $|V_{cb}|$ suggested by eq.~(\ref{eq:Vcb}).
Increasing $|V_{cb}|$ brings the hyperbolic band down more rapidly than
it shrinks the circular band.

If the real-world values of $|V_{cb}|$ and $|V_{ub}/V_{cb}|$ allow a
sizable region, as for the dotted circles in Fig.~\ref{fig:vtd5},
neutral $B$ mixing becomes crucial.
The constraint becomes more restrictive as $x_s$ increases.
Unfortunately, the experimental measurement becomes more difficut as
$x_s$ increases.
If it proves impossible to obtain useful information on $x_s$, one can
return to eq~(\ref{eq:xd}), and focus on $x_d$ alone.
The lattice-QCD calculations of
$\mbox{\small$\frac{8}{3}$}m_B^2f_B^2\hat{B}_B$ will carry larger
uncertainties, however, than the $B_d:B_s$ ratio.

\section{Conclusions}
This talk has examined several ways in which lattice QCD can aid the
determination of standard-model couplings.
The quenched lattice calculations may be divided into several classes,
according to the maturity of the error analysis and the presumed
reliability of the quenched approximation.
One class consists of the light-hadon and quarkonia spectra
and the $K$--$\bar{K}$ mixing parameter $B_K$.
For them the straightforward uncertainties (statistics, $a$, $L$,
excited states, perturbation theory) seem fairly estimated.
The quenched approximation is another matter.
In quarkonia, one can correct for it with potential models, yielding
determinations of $\alpha_S$ and the charm and bottom masses.
The quenched error in $B_K$ is also thought to be under control,
and---taking the error bars at face value---$B_K$ is no longer the
limiting factor in the $|\varepsilon_K|$ constraint on the unitarity
triangle.
A second class consists of $f_B$, the semi-leptonic form factors of $K$
and $D$ mesons (not discussed in this talk, but see \refcite{Som95}),
and the Isgur-Wise function.
These quantities are essential for direct determinations of the first
two rows of the CKM matrix.
The quenched-approximation calculations are in good shape, but the
the corrections to it cannot be simply estimated.
A third class consists of the semi-leptonic decay $B\to\pi l\nu$
and neutral $B$ mixing, for which only exploratory work has appeared.

Nevertheless, all QCD quantities discussed here will follow a
conceptually clear path to ever-more-precise results.
The next ten years or so will almost certainly witness computing and
other technical improvements that will allow for wide-ranging
calculations without the quenched approximation.
By then the most efficient techniques for extracted the most relevant
information will have been perfected.

\section*{Acknowledgements}
The Adriatic glistened in the moonlight as it lapped against the quay.
In the bar of the Hotel Neptun a winsome lounge singer cooed
``\ldots strangers in the night, exchanging glances,\ldots.''
I looked up at the waiter and said, ``Molim pivo,'' when a man strolled
into the bar, clapped me on the back, and cried, ``Ay, you kook, what's
new?''
In a different place and a different time he had rescued my sanity,
if not my life.
I peered into his eyes and replied, ``It's the same old story, always
the same.
And even when it changes, it always ends the same way:
Fermilab is operated by Universities Research Association, Inc.,
under contract DE-AC02-76CH03000 with the U.S. Department of Energy.''

\end{document}